\documentclass[a4paper,fleqn,usenatbib]{mnras}
\usepackage{newtxtext,newtxmath}
\usepackage[T1]{fontenc}
\usepackage{ae,aecompl}

\usepackage{graphicx}
\usepackage{pdflscape} 

\newcommand{\Teff}{$T_{\rm {eff}}$ }
\newcommand{\Msun}{$M_{\odot}$}
\newcommand{\logche}{$\log\,{\textrm {C/He}}$ }
\newcommand{\Ci}{C$\,{\textsc{i}}$ }
\newcommand{\Cii}{C$\,{\textsc{ii}}$ }

\title[The evolution of cool DQ/DQpec white dwarfs]{The evolution of carbon-polluted white dwarfs at low effective temperatures}

\author[S. Blouin \& P. Dufour]{
Simon Blouin$^{1}$\thanks{E-mail: sblouin@lanl.gov} and Patrick Dufour$^{2}$
\\
$^{1}$Los Alamos National Laboratory, P.O. Box 1663, Mail Stop B265, Los Alamos, NM 87545, USA\\
$^{2}$D\'epartement de Physique, Universit\'e de Montr\'eal, Montr\'eal, QC H3C 3J7, Canada\\
}

\pubyear{2019}

\begin{document}
\label{firstpage}
\pagerange{\pageref{firstpage}--\pageref{lastpage}}
\maketitle

\begin{abstract}
Taking advantage of the {\it Gaia} Data Release 2, recent studies have revisited the evolution of carbon-polluted
white dwarfs (DQs) across a large range of effective temperatures. These analyses have clearly confirmed
the existence of two distinct DQ evolutionary sequences: one with normal-mass white dwarfs and one with heavily
polluted and generally more massive objects. The first sequence is thought to result from the dredge-up of carbon from the
core, while the second could at least partially be made of descendants of Hot DQs.
However, the evolution of carbon-polluted white dwarfs below 6500\,K remains unexplored, mainly due
to the theoretical difficulties associated with modelling their dense atmospheres.
In this work, we present a detailed star-by-star analysis of cool carbon-polluted white dwarfs.
Our recently improved atmosphere models allow us to obtain good fits to most objects, including very
cool DQpec white dwarfs with strongly shifted C$_2$ molecular bands.
We show that cool carbon-polluted white dwarfs keep following the two distinct evolutionary tracks previously
identified at higher temperatures. We also find that most DQ white dwarfs transform into DQpec
when their photospheric densities exceed $\approx 0.15\,{\textrm{g\,cm}}^{-3}$.
However, we identify stars for which the DQ$\rightarrow$DQpec transition occurs at lower
photospheric densities, possibly due to the presence of a strong magnetic field.
\end{abstract}

\begin{keywords}
stars: abundances -- stars: evolution -- white dwarfs
\end{keywords}



\section{Introduction}
A DQ star is a white dwarf whose spectrum is dominated by carbon features. Depending on the effective
temperature of a DQ white dwarf, the spectroscopic signature of carbon can take different forms. For warm DQ
white dwarfs ($T_{\rm eff} \approx 10{,}000 - 16{,}000\,{\rm K}$), carbon is detected as \Ci
atomic lines. At lower temperatures ($T_{\rm eff} < 10{,}000\,{\rm K}$), the formation of the C$_2$ 
molecule leads to the appearance of strong molecular bands, the most prominent being the Swan
bands ($4500-6000\,{\AA}$). Spectroscopic analyses of cool DQ white dwarfs using atmosphere models
have shown that the atmosphere of those objects is dominated by helium and polluted by small
quantities of carbon \citep[$-7 < \log\,{\rm C/He} < -2$, e.g.,][]{koester1982,weidemann1995,dufour2005,koester2006}.
The presence of carbon in the atmosphere of DQ stars implies the existence of a mechanism that
thwarts the efficient gravitational settling at play in white dwarfs. For most DQs, this mechanism
is the transport of carbon from the deep interior by the helium convection zone \citep{pelletier1986,fontaine2005}.
In fact, for the bulk of objects, the observed monotonic decrease of \logche with decreasing \Teff
\citep{dufour2005,koester2006,kepler2016} is well accounted for by this model \citep[e.g.,][Figure 12]{coutu2019}.

However, the dredge-up model fails to explain the observed composition of a second sequence of DQ
white dwarfs that, for a given $T_{\rm eff}$, have carbon abundances about one order of magnitude higher
than the main DQ sequence \citep[e.g.,][Figure 12]{coutu2019}. As clearly
revealed by the {\it Gaia} DR2 parallaxes, above $T_{\rm eff} \approx 10{,}000\,{\rm K}$,
this second sequence is made of more massive objects 
\citep[$\langle M \rangle \approx 1\,M_{\odot}$,][]{coutu2019,koester2019}. The problem is that high-mass
evolutionary sequences simply cannot match the slope of the observed decrease of \logche with decreasing \Teff
\citep[][Figure 1]{brassard2007}. The fact that evolutionary models fail to account for the
atmospheric composition of those objects---while successfully accounting for the composition of DQs
on the first sequence---clearly suggests that another scenario than the dredge-up model must be invoked
to explain carbon pollution in the second sequence. The preferred scenario
is that they are the product of the evolution of another white dwarf spectral type, Hot DQs \citep{coutu2019}.
The spectra of Hot DQ white dwarfs are characterized by \Ci and \Cii atomic lines \citep{liebert2003}.
Model atmosphere analyses have shown that those stars have carbon-dominated atmospheres and effective
temperatures ranging from $18{,}000$ to $24{,}000\,{\rm K}$ \citep{dufour2007hotdq,dufour2008}.
Hot DQ white dwarfs are thought to originate from merged white dwarfs \citep{dunlapphd,dunlap2015}. This
hypothesis is supported by the high velocity dispersion of those objects, which suggests that
they are much older than the age inferred from their atmospheric parameters. 
As a merger event would lead to a significant reheating, the cooling age derived from the temperature
becomes meaningless, which naturally explains the mismatch with the kinematic age.
Many pieces of evidence support the idea that, at least above $T_{\rm eff} \approx 10{,}000\,{\rm K}$,
the second DQ sequence is made of the descendants of Hot DQs:
(1) the second sequence connects nicely with the Hot DQs in a $T_{\rm eff}-\log\,{\rm C/He}$ 
diagram \citep{dufour2013,coutu2019}, thus suggesting a common evolutionary origin;
(2) DQs from the second sequence have similar kinematic properties as Hot DQs \citep{dunlap2015,coutu2019}\footnote{We note, however, that the recent results of \cite{cheng2019} indicate that a merger time delay 
alone cannot not fully explain the kinematic ages of the massive DQs of the second sequence.};
(3) both populations are characterized by high masses \citep[Dunlap et al. submitted,][]{coutu2019}.

The very recent studies of \cite{coutu2019} and \cite{koester2019} represent an important step forward
in our understanding of the evolution of carbon-polluted white dwarfs. However, their analyses stop
short of investigating the evolution of DQ white dwarfs at very cool effective temperatures
($T_{\rm eff}<6500\,{\rm K}$). Cool, helium-rich white dwarf atmospheres are characterized by high
photospheric densities \citep[e.g.,][Figure 13]{blouin2017}. Under such conditions, the radiative opacities,
the equation of state and the chemical equilibrium can significantly differ from the ideal gas
results \citep{kowalski2006,kowalski2007,blouin2018a,rohrmann2018}. In particular, the C$_2$ Swan bands
undergo a density-driven distortion that has been attributed to a shift of the electronic transition
energy \citep{kowalski2010}. White dwarfs with such distorted Swan bands are known as DQpec stars
\citep{hall2008}. For a lack of atmosphere models accounting for those high-density effects,
\cite{coutu2019} have ignored all DQpec white dwarfs from their analysis as well as all DQs with
an effective temperature below 6000\,K. Regarding the study of \cite{koester2019}, a crude analysis
of DQpec white dwarfs was performed assuming a constant effective temperature and carbon abundance for all objects.
Based on this analysis, they concluded that DQpec white dwarfs might be massive objects that are
the descendants of the Hot DQs.

The aim of this work is to establish the evolution of DQ white dwarfs at very cool effective temperatures
($T_{\rm eff}<6500\,{\rm K}$). This is done using our recently improved atmosphere models and a sample
of all known cool DQ white dwarfs. Our models
and the selection of our sample are described in Section~\ref{sec:methodology}. We present our model
atmosphere analysis in Section~\ref{sec:results} and the implications of our results on the evolution
of carbon-polluted white dwarfs in Section~\ref{sec:evolution}. Finally, our main conclusions are
given in Section~\ref{sec:conclu}.

\section{Methodology}
\label{sec:methodology}
\subsection{Atmosphere models}
The atmosphere code used in this work is identical to that described in \cite{blouin2019c}. This code
is uniquely suited for the study of cool helium-rich white dwarfs as
it includes an accurate description of the effects of a high helium density on the chemical equilibrium
and on the radiative opacities \citep[][and references therein]{blouin2018a,blouin2018b}.
The opacity of the C$_2$ Swan bands is computed with a line-by-line approach that uses a linelist
provided by J.~O. Hornkohl (private communication), which was obtained following the methodology
described in \cite{parigger2015}. Moreover, following the work of \cite{kowalski2010},
we include a density-driven shift of the electronic transition energy of the Swan bands.
This shift is computed as $\Delta T_e (\rm eV) = \alpha \rho (\textrm{g\,cm}^{-3})$, where $\alpha=0.2$
as empirically determined in \citet[Section 3.3]{blouin2019c}.\footnote{The $\alpha=0.2$ value differs
significantly from the $\alpha=1.6$ value obtained from density functional theory calculations. See
\citet[Section 3.3]{blouin2019c} for a detailed discussion of this problem.}
We use a 3-dimensional grid of model atmospheres, with \Teff varying from 4000\,K to 9000\,K in steps of
500\,K, $\log g$ from 7.0 to 9.0 in steps of 0.5 dex, and \logche from $-9.0$ to $-4.0$ in steps of 0.5 dex.
Note that our models do not include any hydrogen. This is justified by the finding that for $T_{\rm eff}<8000\,{\rm K}$
the CH $G$ band should be visible even for hydrogen abundances that lead to a negligible impact on the model
and on the derived atmospheric parameters \citep{blouin2019c,coutu2019}. Only one object in our sample, 
G99$-$37 (GJ~1086), displays a CH $G$ band. For this object, we rely on the solution already provided in \cite{blouin2019c}.

One major caveat of our models is the current uncertainty surrounding ultraviolet opacities. The carbon atomic lines are 
included using the Vienna Atomic Line Database \citep[VALD,][]{vald1,vald2,vald3}. However, as discussed in \cite{coutu2019}
and \cite{koester2019}, many carbon lines in the ultraviolet are predicted to be much stronger and wider than
they appear in observed spectra. This problem is probably at the origin of a $\approx 0.05\,M_{\odot}$ shift 
of the peak of the DQ mass distribution with respect to that of DA and DB white dwarfs \citep{coutu2019}.
This finding casts some doubts on the absolute values of the derived atmospheric parameters, but the effect on
the relative values between objects is expected to be minimal.

\begin{landscape}
  \begin{table}
    \scriptsize
    \centering
    \caption{Observational data.}
    \label{tab:obs}
\begin{tabular}{llccccccccccrrcc}
\hline
               SDSS J &                     MWDD ID &    \multicolumn{5}{c}{SDSS} &  \multicolumn{5}{c}{Pan--STARRS} &  \multicolumn{1}{c}{$\pi$} &  \multicolumn{1}{c}{$\sigma_{\pi}$}  &       Spectrum & Magnetic? \\
               &                     &   $u$ &     $g$ &     $r$ &     $i$ &     $z$ &    $g$ &   $r$ &    $i$ &    $z$ &   $y$ &     (mas) & (mas) & source  & \\

\hline
                   -- &                     GJ 2012 &     -- &     -- &     -- &     -- &     -- &  14.76 &  14.29 &  14.11 &  14.06 &  14.03 &  109.88 &  0.03 &  (1) &  No ($<3\,{\rm MG}$)$^a$        \\
                   -- &                   LP 410$-$80 &     -- &     -- &     -- &     -- &     -- &  17.17 &  16.91 &  16.91 &  16.93 &  16.99 &   23.71 &  0.09 &        (2) &           \\
                   -- &                    Wolf 219 &     -- &     -- &     -- &     -- &     -- &  15.28 &  15.08 &  15.06 &  15.15 &  15.14 &   53.00 &  0.05 &  (1) &    No$^b$     \\
                   -- &                    LP 717$-$1 &     -- &     -- &     -- &     -- &     -- &  17.55 &  17.04 &  16.88 &  16.86 &  16.83 &   28.83 &  0.08 &        (2) &           \\
                   -- &                     GJ 1086 &     -- &     -- &     -- &     -- &     -- &  14.77 &  14.40 &  14.35 &  14.39 &  14.44 &   89.17 &  0.03 &  (1) &     $\approx 7.5\,{\rm MG}^c$      \\
   080455.42+171443.6 &    SDSS J080455.42+171443.6 &  19.92 &  19.05 &  18.46 &  18.26 &  18.23 &  18.99 &  18.44 &  18.29 &  18.25 &  18.22 &   14.95 &  0.33 &             SDSS &           \\
   080558.84+072448.5 &    SDSS J080558.83+072447.8 &  20.45 &  19.55 &  18.89 &  18.69 &  18.66 &  19.48 &  18.89 &  18.72 &  18.66 &  18.72 &   12.80 &  0.35 &             SDSS &           \\
   080843.15+464028.6 &                 WD 0805+468 &  20.69 &  20.37 &  19.35 &  19.10 &  19.03 &  20.40 &  19.42 &  19.13 &  19.15 &  19.23 &    9.67 &  0.72 &             SDSS &           \\
   082955.77+183532.6 &  [VV2010c] J082955.8+183532 &  22.74 &  21.82 &  20.45 &  20.17 &  20.18 &  21.68 &  20.46 &  20.20 &  20.03 &  20.02 &    5.74 &  1.75 &             SDSS &           \\
   083618.13+243254.6 &    SDSS J083618.13+243254.6 &  20.30 &  19.51 &  18.90 &  18.74 &  18.70 &  19.48 &  18.94 &  18.73 &  18.75 &  18.65 &   10.23 &  0.93 &             SDSS &           \\
   090208.40+201049.9 &                   LP 426-49 &  18.95 &  18.87 &  17.79 &  17.25 &  17.29 &  18.90 &  17.75 &  17.26 &   17.30 &  17.36 &   25.92 &  0.17 &             SDSS &           \\
   090632.17+470235.8 &    SDSS J090632.17+470235.8 &  20.58 &  20.36 &  19.44 &  19.02 &  19.17 &  20.33 &  19.42 &  19.03 &  19.07 &  19.08 &   11.65 &  0.42 &             SDSS &           \\
   093537.00+002422.0 &                 WD 0933+006 &  20.28 &  20.15 &  19.18 &  18.66 &  18.66 &  20.08 &  19.15 &  18.63 &   18.70 &  18.71 &   12.89 &  0.38 &             SDSS &           \\
   101141.53+284556.0 &                   LP 315$-$42 &  18.27 &  18.24 &  16.42 &  15.97 &  15.99 &  18.26 &  16.40 &  15.97 &     16.00 &  16.01 &   67.79 &  0.08 &             SDSS &    $\sim 100\,{\rm MG}^d$       \\
                   -- &                     GJ 3614 &     -- &     -- &     -- &     -- &     -- &  16.62 &  15.75 &  15.72 &  15.71 &  15.65 &   70.76 &  0.07 &  (1) &      $50-200\,{\rm MG}^e$     \\
                   -- &                 BD$-$18 3019B &     -- &     -- &     -- &     -- &     -- &     -- &     -- &     -- &     -- &     -- &   53.13 &  0.06 &  (1) &           \\
   111341.33+014641.7 &                 WD 1111+020 &  18.66 &  19.19 &  18.47 &  18.28 &  18.10 &  19.25 &  18.47 &  18.25 &  18.12 &  18.06 &   22.95 &  0.24 &             SDSS &    Yes$^f$       \\
   112036.74+010629.3 &    SDSS J112036.74+010629.3 &  21.25 &  20.93 &  20.23 &  20.07 &  20.17 &  20.98 &  20.24 &  20.08 &  20.09 &  19.93 &    3.94 &  1.08 &             SDSS &           \\
   115933.10+130031.6 &                 WD 1156+132 &  18.24 &  18.14 &  17.75 &  17.67 &  17.78 &  18.19 &  17.76 &  17.70 &  17.81 &  17.91 &   16.10 &  0.22 &             SDSS &           \\
   121037.44+140644.4 &    SDSS J121037.44+140644.4 &  21.54 &  20.61 &  20.03 &  19.85 &  19.79 &  20.55 &  19.99 &  19.86 &  19.81 &  19.72 &    7.80 &  0.84 &             SDSS &           \\
   122545.88+470613.0 &       PSO J186.4406+47.1036 &  19.77 &  19.58 &  19.08 &  18.89 &  18.96 &  19.60 &  19.08 &  18.89 &  19.01 &  19.02 &    8.88 &  0.32 &             SDSS &           \\
   123313.48+082403.1 &                  NLTT 31076 &  19.15 &  18.64 &  18.35 &  18.30 &  18.35 &  18.62 &  18.37 &  18.31 &  18.35 &   18.40 &   12.25 &  0.29 &             SDSS &           \\
   124733.69+491524.7 &    SDSS J124733.70+491524.8 &  19.94 &  19.39 &  19.04 &  18.95 &  18.96 &  19.34 &  19.05 &  18.96 &  18.97 &  18.96 &    9.11 &  0.30 &             SDSS &           \\
   124739.05+064604.6 &              PM J12476+0646 &  20.95 &  20.03 &  18.68 &  18.39 &  18.27 &  19.90 &  18.69 &  18.41 &  18.35 &  18.31 &   19.06 &  0.28 &      (3) &           \\
   131146.93+292351.1 &                 WD 1309+296 &  19.66 &  19.44 &  18.50 &  17.93 &  17.98 &  19.52 &  18.44 &  17.91 &  17.95 &  17.95 &   18.42 &  0.19 &             SDSS &           \\
   133359.86+001654.8 &                 WD 1331+005 &  19.11 &  19.40 &  18.37 &  18.18 &  18.24 &  19.50 &  18.36 &  18.12 &   18.20 &  18.21 &   24.45 &  0.35 &             SDSS &   Yes$^f$        \\
   134118.68+022736.9 &                 WD 1338+027 &  18.38 &  17.96 &  17.29 &  17.17 &  17.18 &  17.93 &  17.32 &  17.17 &  17.18 &  17.18 &   24.88 &  0.13 &             SDSS &           \\
   145725.27+210747.3 &    SDSS J145725.27+210747.3 &  19.71 &  19.13 &  18.68 &  18.55 &  18.53 &  19.08 &  18.68 &  18.58 &  18.56 &  18.57 &   11.99 &  0.28 &             SDSS &           \\
   161140.18+045127.0 &     USNO$-$B1.0 0948$-$00255808 &  19.45 &  18.69 &  18.27 &  18.13 &  18.08 &  18.64 &  18.27 &  18.16 &  18.12 &  18.18 &   14.82 &  0.18 &             SDSS &           \\
   161414.12+172900.5 &                   LP 444$-$33 &  19.20 &  18.66 &  17.86 &  17.74 &  17.81 &  18.65 &  17.88 &  17.75 &  17.81 &  17.93 &   18.43 &  0.13 &             SDSS &           \\
   161847.38+061155.2 &    SDSS J161847.38+061155.2 &  18.41 &  18.23 &  18.26 &  18.45 &  18.43 &  18.27 &  18.25 &  18.45 &  18.52 &   18.50 &   12.94 &  0.19 &             SDSS &           \\
   162635.58+154441.6 &                          -- &  20.23 &  19.66 &  19.40 &  19.40 &  19.32 &  19.67 &  19.38 &  19.38 &  19.44 &  19.47 &    7.98 &  0.36 &             SDSS &           \\
   180302.57+232043.3 &    SDSS J180302.57+232043.3 &  21.41 &  20.44 &  19.02 &  18.74 &  18.63 &     -- &     -- &     -- &     -- &     -- &   15.33 &  0.32 &             SDSS &           \\
   183500.21+642917.0 &    SDSS J183500.21+642917.0 &  17.68 &  17.59 &  17.21 &  17.14 &  17.27 &  17.65 &  17.23 &  17.17 &  17.28 &  17.39 &   18.73 &  0.07 &             SDSS &           \\
  223224.00$-$074434.3 &                 WD 2229$-$080 &  18.48 &  18.40 &  17.78 &  17.66 &  17.75 &  18.43 &  17.80 &  17.68 &  17.79 &  17.87 &   18.27 &  0.17 &             SDSS &           \\
   224153.46+043256.6 &                  NLTT 54596 &  18.99 &  18.27 &  17.92 &  17.80 &  17.82 &  18.29 &  17.94 &  17.84 &  17.83 &  17.89 &   16.20 &  0.23 &      (3) &           \\
   225901.16+215843.9 &                          -- &  20.65 &  20.75 &  20.13 &  19.96 &  20.35 &  20.72 &  20.08 &  19.94 &  20.07 &  19.91 &    6.80 &  0.84 &             SDSS &           \\

\hline
\multicolumn{16}{l}{Note: for BD$-$18 3019B, we use the $BVR$ and $JHK$ photometry from \cite{bergeron2001}.}\\
\multicolumn{16}{l}{(1) \cite{giammichele2012}; (2) \cite{kawka2012}; (3) \cite{limoges2015}}\\
\multicolumn{16}{l}{$^a$\cite{schmidt1995}; $^b$\cite{vornanen2013}; $^c$\cite{berdyugina2007}; $^d$\cite{schmidt1999}; 
$^e$\cite{jordan2002}; $^f$\cite{schmidt2003}}
\end{tabular}
\end{table}
\end{landscape}

\subsection{Observational data}
Our sample is made of all known cool ($T_{\rm eff}<7000{\rm K}$) carbon-polluted white dwarfs.
To identify which objects
to include in our sample, we relied on the Montreal White Dwarf Database \citep[MWDD,][]{mwdd} and on the
sample of \cite{koester2019}.
Objects already fitted in \cite{blouin2019c} or \cite{coutu2019} are not analysed in Section~\ref{sec:results},
since these fits were performed using the same atmosphere code (i.e., same input physics)
as the one used in the present work. Nevertheless, those stars are included in our study of the evolution of
cool carbon-polluted white dwarfs (Section~\ref{sec:evolution}).

For every object in our sample, we have retrieved the {\it Gaia} DR2 parallax measurement \citep{gaiadr2a,gaiadr2b},
as well as the $ugriz$ photometry from the Sloan Digital Sky Survey \citep[SDSS,][]{sdss}
and the $grizy$ photometry from
the Panoramic Survey Telescope and Rapid Response System \citep[Pan--STARRS,][]{panstarrs} when available.
Table \ref{tab:obs} lists the observational data for the 37 objects included in our sample. Also shown is the source
of the spectrum used to derive the C/He abundance ratio as well as the available information about the 
presence of magnetic fields at the surface of those objects.

\section{Results}
\label{sec:results}
To extract the atmospheric parameters from the photometric and spectroscopic data, we follow the procedure
described in \cite{dufour2005}. The first step consists of finding \Teff and the solid angle $\pi (R/D)^2$
by fitting the synthetic photometry to the photometric observations (we use both the $ugriz$ and $grizy$
photometry when an object is in SDSS and Pan--STARRS). Since the distance $D$ is already
known from the {\it Gaia} parallax, the radius $R$ can be obtained. From there, we compute the mass $M$
and the surface gravity $\log g$ using evolutionary models similar to those described in \cite{fontaine2001},
assuming C/O cores and $\log M_{\rm He}/M_{\star} = -2$.
Then, while keeping \Teff and $\log g$ constant, we determine \logche by adjusting the synthetic spectrum
to the observed Swan bands. The C/He ratio thus found is usually different from the abundance initially
assumed for the photometric fit. We therefore repeat this procedure (i.e., the photometric and the
spectroscopic fits) until $T_{\rm eff}$, $\log g$ and \logche converge to a definitive solution.

\begin{figure}
  \centering
  \includegraphics[width=\columnwidth]{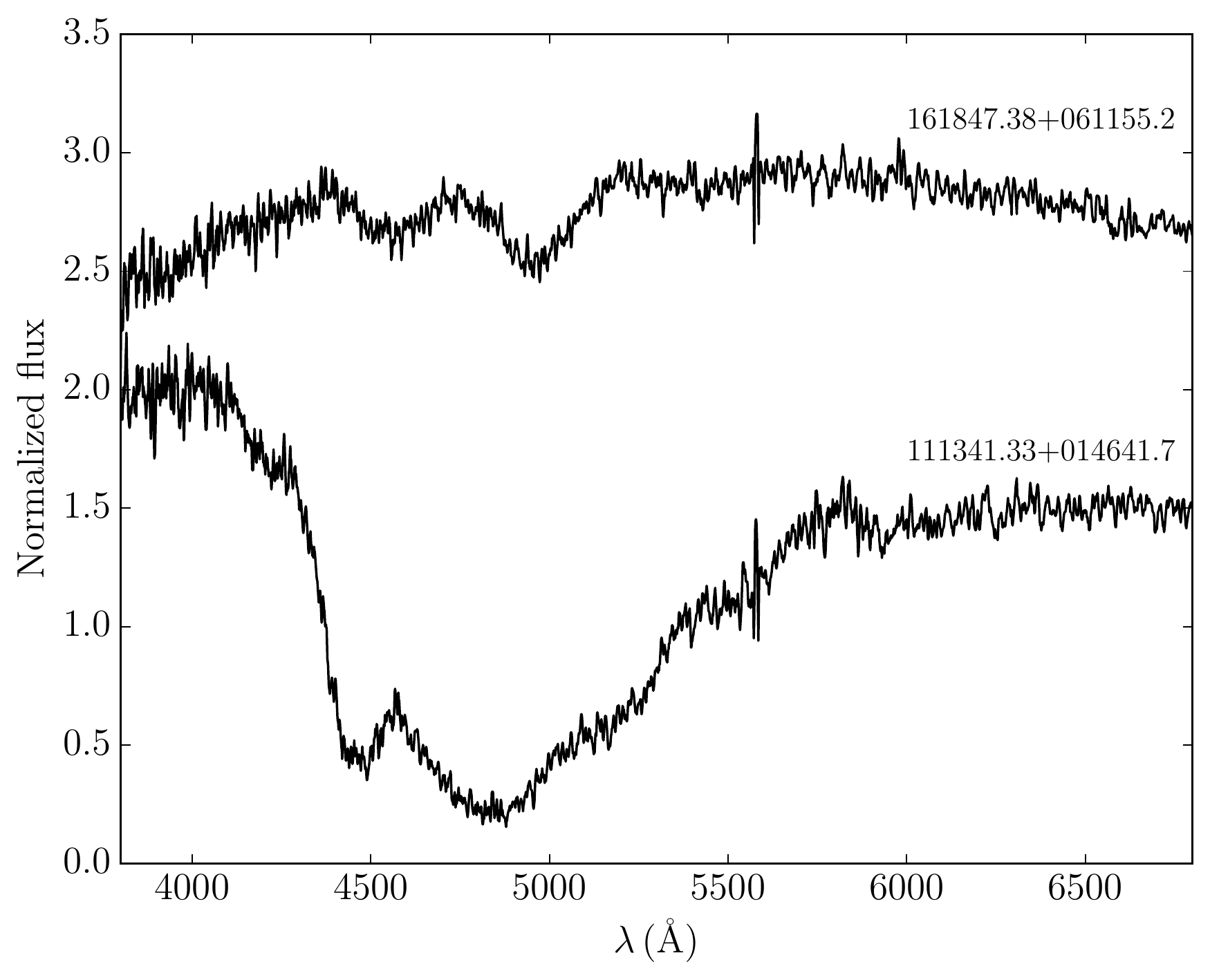}
  \caption{SDSS spectra of two unusual objects. Both spectra were smoothed and the spectrum of
    SDSS J161847.38+061155.2 was vertically shifted for clarity.}
  \label{fig:weird}
\end{figure}

In the case of SDSS J111341.33+014641.7, we did not follow the procedure outlined above. This star shows
extremely strong and atypically distorted molecular bands (Figure~\ref{fig:weird}) that our models
are completely unable to reproduce. Therefore, we estimated the C/He ratio by manually adjusting
\logche to match the depth of the bands. Obviously, this approximative procedure
leads to high uncertainties on the atmospheric parameters.

\begin{figure*}
  \centering
  \includegraphics[width=\textwidth]{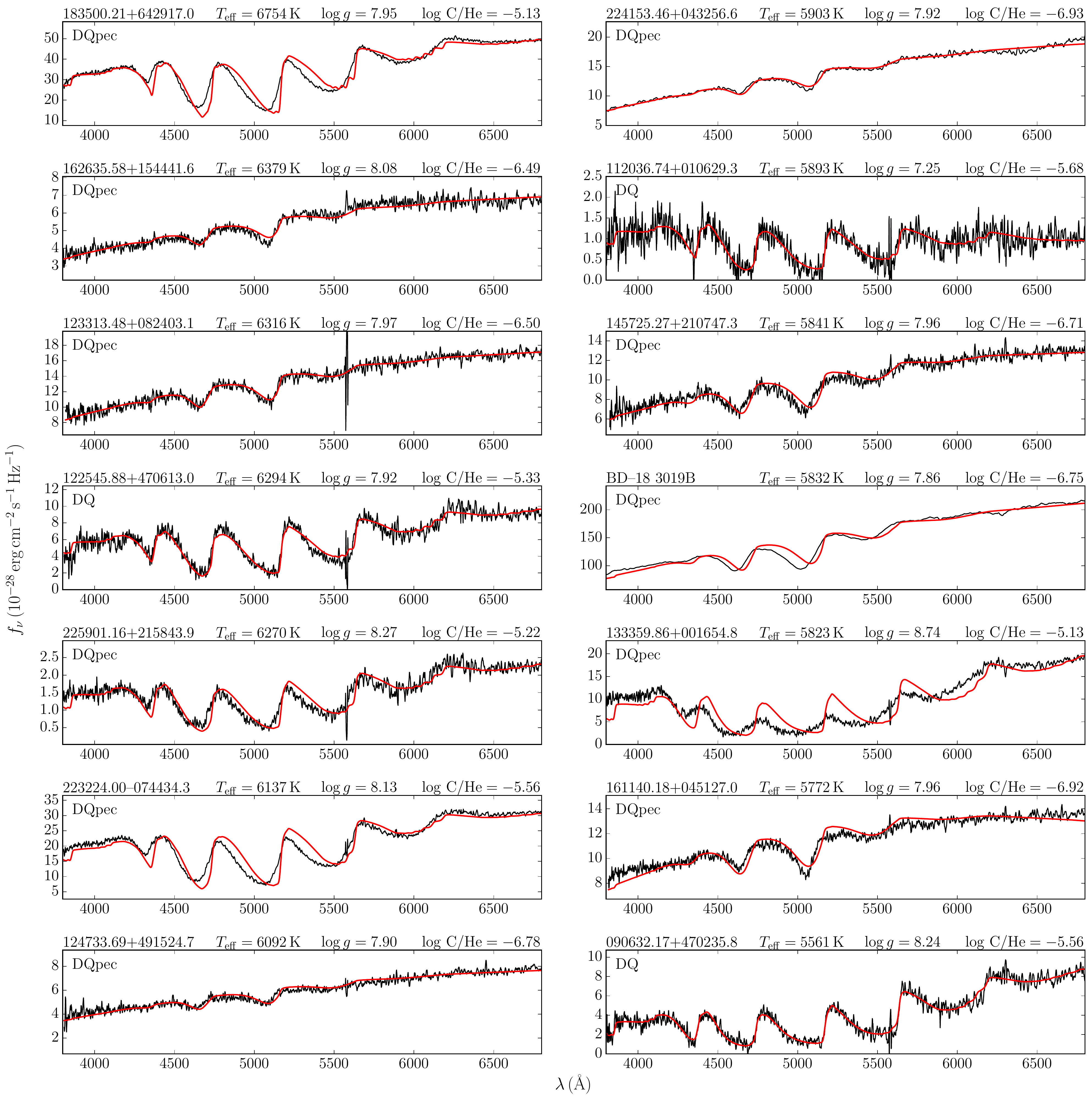}
  \caption{Spectroscopic fits of DQ/DQpec objects in our sample. Objects are shown in order of decreasing $T_{\rm eff}$.
    The DQ/DQpec classification is described in Section \ref{sec:transition}.}
  \label{fig:fits}
\end{figure*}

Table~\ref{tab:sol} lists the atmospheric parameters found following our fitting procedure
(SDSS J161847.38+061155.2 is omitted from this table, see Section \ref{sec:problems}) and 
Figure~\ref{fig:fits} displays our fit to the spectroscopic data. Note that stars that were already
analysed in \cite{blouin2019c} are not shown in Figure~\ref{fig:fits} as we assumed the same atmospheric
parameters as those determined in that work (since our atmosphere code is unchanged).
Most of the fits presented in Figure \ref{fig:fits} are satisfactory. The shape and position of the Swan bands
are generally well reproduced, even for the most extreme objects. In particular, our model matches the very 
deep absorption bands of SDSS J131146.93+292351.1 (GSC2U J131147.2+292348) much more closely than what was
obtained with previous atmosphere codes \citep{carollo2003,dufour2005}. Moreover, we achieve a good fit for the most extreme
DQpec white dwarfs of our sample (i.e., SDSS J124739.05+064604.6, SDSS J082955.77+183532.6, and
SDSS J180302.57+232043, the three coolest star in Figure~\ref{fig:fits}). 
Our models successfully reproduce the pronounced shift of the Swan bands for the
high densities encountered at the photosphere of those objects ($\rho \approx 0.7\,$g\,cm$^{-3}$).

\begin{table*}
  \centering
  \caption{Atmospheric parameters.}
  \label{tab:sol}
  \begin{tabular}{lllrrcl}
\hline
              SDSS J &                     MWDD ID &  \multicolumn{1}{c}{\Teff}  &  \multicolumn{1}{c}{$\log g$}  &      \multicolumn{1}{c}{$M$} &  \multicolumn{1}{c}{\logche$^a$} & Spectral \\
                     &                             &  \multicolumn{1}{c}{(K)}           &                 &     \multicolumn{1}{c}{(\Msun)}& (by number)  & \multicolumn{1}{c}{type}    \\
\hline
                  -- &                     GJ 2012 &  5210   (60) &  7.907 (0.038)  &  0.514 (0.023)  &   $-$8.40  & DQpec\\
                  -- &                 LP 410$-$80 &  6360   (60) &  7.993 (0.022)  &  0.569 (0.013)  &   $-$6.47  & DQpec\\
                  -- &                    Wolf 219 &  6515   (60) &  7.974 (0.025)  &  0.558 (0.015)  &   $-$6.45  & DQpec\\
                  -- &                  LP 717$-$1 &  5375   (20) &  7.906 (0.014)  &  0.514 (0.008)  &   $-$7.41  & DQpec\\
                  -- &                     GJ 1086 &  6080   (45) &  8.146 (0.019)  &  0.663 (0.012)  &   $-$6.57  & DQ   \\
  080455.42+171443.6 &    SDSS J080455.42+171443.6 &  5364   (33) &  7.885 (0.038)  &  0.502 (0.022)  &   $-$7.19  & DQpec\\
  080558.84+072448.5 &    SDSS J080558.83+072447.8 &  5313   (42) &  7.938 (0.046)  &  0.532 (0.027)  &   $-$7.09  & DQpec\\
  080843.15+464028.6 &                 WD 0805+468 &  5427   (64) &  7.913 (0.128)  &  0.518 (0.073)  &   $-$5.93  & DQpec\\
  082955.77+183532.6 &  [VV2010c] J082955.8+183532 &  4500   (24) &  7.113 (0.787)  &  0.188 (0.320)  &   $-$7.76  & DQpec\\
  083618.13+243254.6 &    SDSS J083618.13+243254.6 &  5277   (54) &  7.545 (0.174)  &  0.335 (0.072)  &   $-$7.39  & DQpec\\
  090208.40+201049.9 &                 LP 426$-$49 &  5499  (110) &  8.224 (0.021)  &  0.713 (0.014)  &   $-$5.37  & DQ   \\
  090632.17+470235.8 &    SDSS J090632.17+470235.8 &  5561   (68) &  8.238 (0.053)  &  0.722 (0.035)  &   $-$5.56  & DQ   \\
  093537.00+002422.0 &                 WD 0933+006 &  5539   (60) &  8.144 (0.046)  &  0.660 (0.020)  &   $-$5.37  & DQ   \\
  101141.53+284556.0 &                 LP 315$-$42 &  4335  (165) &  8.211 (0.085)  &  0.703 (0.057)  &   $-$6.80  & DQpec\\
                  -- &                     GJ 3614 &  4530  (215) &  8.074 (0.124)  &  0.614 (0.078)  &   $-$7.20  & DQpec\\
                  -- &               BD$-$18 3019B &  5832   (86) &  7.863 (0.052)  &  0.492 (0.029)  &   $-$6.74  & DQpec\\
  111341.33+014641.7 &                 WD 1111+020 &  5961  (350) &  8.709 (0.122)  &  1.027 (0.073)  &   $-$5.14  & DQpec\\
  112036.74+010629.3 &    SDSS J112036.74+010629.3 &  5893   (40) &  7.253 (0.640)  &  0.320 (0.180)  &   $-$5.68  & DQ   \\
  115933.10+130031.6 &                 WD 1156+132 &  6410   (60) &  8.011 (0.034)  &  0.580 (0.021)  &   $-$5.70  & DQ   \\
  121037.44+140644.4 &    SDSS J121037.44+140644.4 &  5419   (23) &  8.031 (0.174)  &  0.589 (0.107)  &   $-$7.00  & DQpec\\
  122545.88+470613.0 &       PSO J186.4406+47.1036 &  6294   (67) &  7.924 (0.059)  &  0.528 (0.034)  &   $-$5.33  & DQ   \\
  123313.48+082403.1 &                  NLTT 31076 &  6316   (61) &  7.970 (0.041)  &  0.555 (0.024)  &   $-$6.50  & DQpec\\
  124733.69+491524.7 &    SDSS J124733.70+491524.8 &  6092   (38) &  7.897 (0.058)  &  0.512 (0.033)  &   $-$6.78  & DQpec\\
  124739.05+064604.6 &              PM J12476+0646 &  4545   (31) &  7.929 (0.033)  &  0.526 (0.019)  &   $-$7.50  & DQpec\\
  131146.93+292351.1 &                 WD 1309+296 &  5529   (40) &  8.178 (0.018)  &  0.683 (0.012)  &   $-$5.27  & DQ   \\
  133359.86+001654.8 &                 WD 1331+005 &  5823   (71) &  8.741 (0.038)  &  1.052 (0.022)  &   $-$5.13  & DQpec\\
  134118.68+022736.9 &                 WD 1338+027 &  5785   (20) &  8.098 (0.013)  &  0.632 (0.008)  &   $-$6.00  & DQpec\\
  145725.27+210747.3 &    SDSS J145725.27+210747.3 &  5841   (36) &  7.962 (0.043)  &  0.548 (0.025)  &   $-$6.71  & DQpec\\
  161140.18+045127.0 &   USNO-B1.0 0948$-$00255808 &  5772   (30) &  7.956 (0.026)  &  0.545 (0.015)  &   $-$6.92  & DQpec\\
  161414.12+172900.5 &                 LP 444$-$33 &  5491  (101) &  7.935 (0.059)  &  0.531 (0.034)  &   $-$6.73  & DQpec\\
  162635.58+154441.6 &                          -- &  6379   (51) &  8.082 (0.073)  &  0.623 (0.046)  &   $-$6.49  & DQpec\\
  180302.57+232043.3 &    SDSS J180302.57+232043.3 &  4400   (67) &  7.694 (0.086)  &  0.400 (0.042)  &   $-$7.79  & DQpec\\
  183500.21+642917.0 &    SDSS J183500.21+642917.0 &  6754   (40) &  7.951 (0.018)  &  0.545 (0.010)  &   $-$5.14  & DQpec\\
  223224.00$-$074434.3 &             WD 2229$-$080 &  6137   (44) &  8.126 (0.028)  &  0.651 (0.018)  &   $-$5.56  & DQpec\\
  224153.46+043256.6 &                  NLTT 54596 &  5903   (38) &  7.920 (0.030)  &  0.524 (0.017)  &   $-$6.93  & DQpec\\
  225901.16+215843.9 &                          -- &  6270   (49) &  8.267 (0.180)  &  0.744 (0.119)  &   $-$5.22  & DQpec\\
\hline
\multicolumn{7}{l}{$^a$ Typically, the uncertainy on \logche is 0.10 dex.}
\end{tabular}
\end{table*}

\renewcommand{\thefigure}{\arabic{figure} (cont.)}
\addtocounter{figure}{-1}
\begin{figure*}
  \centering
  \includegraphics[width=\textwidth]{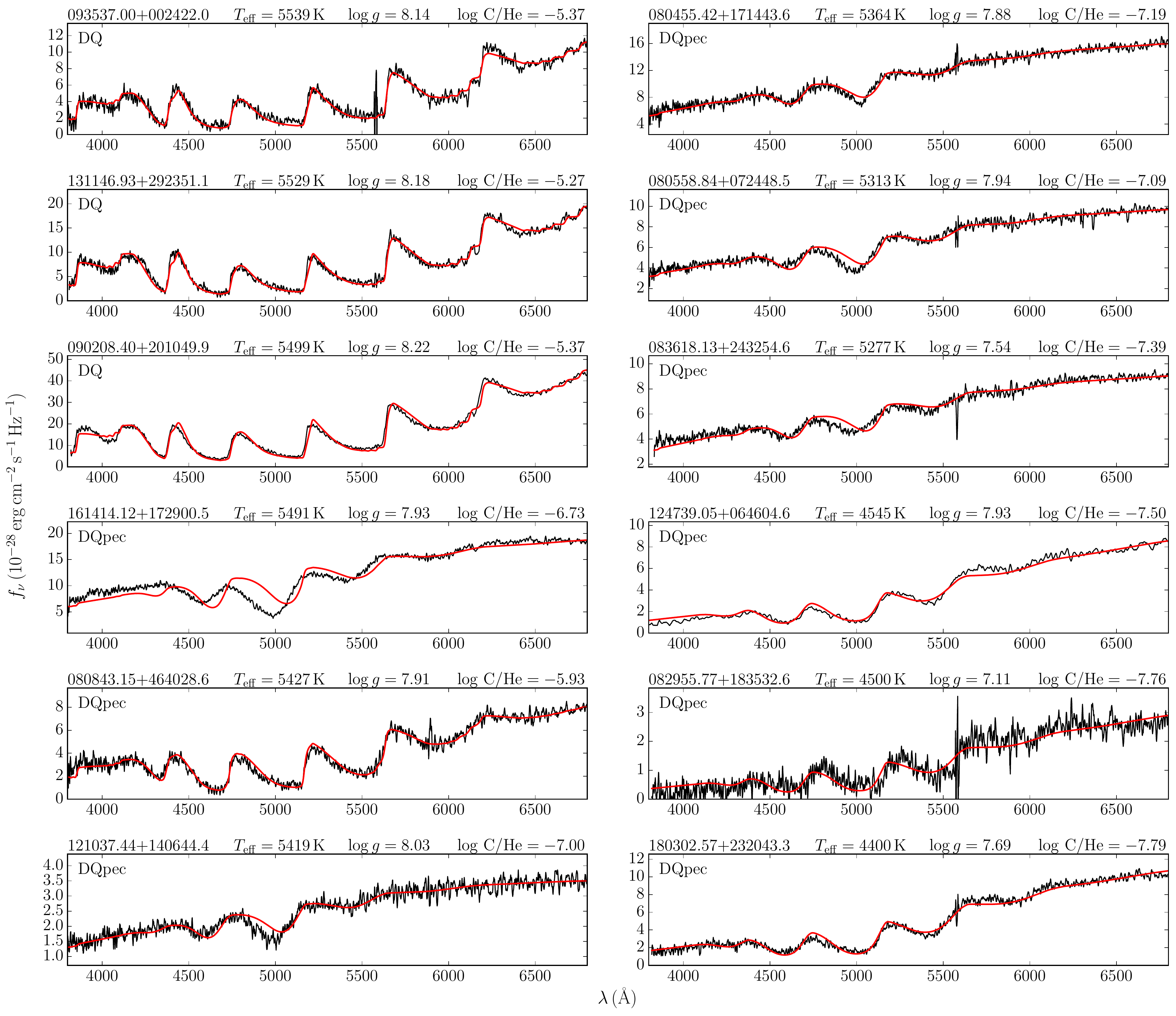}
  \caption{Spectroscopic fits of DQ/DQpec objects in our sample. Objects are shown in order of decreasing $T_{\rm eff}$.
  The DQ/DQpec classification is described in Section \ref{sec:transition}.}
\end{figure*}
\renewcommand{\thefigure}{\arabic{figure}}

\subsection{Problematic objects}
\label{sec:problems}
That being said, there is a rather diverse group of objects
(BD$-$18 3019B, SDSS J111341.33+014641.7, SDSS J133359.86+001654.8, SDSS J161414.12+172900.5,
SDSS J183500.21+642917.0, SDSS J223224.00$-$074434.3, SDSS J225901.16+215843.9)
for which our models
underestimate the amplitude of the Swan bands shift. The problem for those objects is that the 
effective temperature inferred from the
photometric fit is too high to result in a photospheric density that would be high enough to
sufficiently shift the Swan bands. The origin of the problem is unclear since these objects span
a wide range of effective temperatures, carbon abundances and surface gravities. One possibility
is that the empirically determined $\Delta T_e (\rm eV) = 0.2 \rho (\textrm{g\,cm}^{-3})$ shift,
while valid for most objects, is too simplistic to properly capture the distortion of
Swan bands under all physical conditions. Another possibility, also discussed in \cite{blouin2019c}
in the context of other problematic objects, is that those stars may harbour a strong magnetic
field that could affect their structures through the suppression of convection
\citep{tremblay2015,gentile2018} and their radiative opacities through the magnetic distortion of the
C$_2$ Swan bands \citep{liebert1978,bues1991,bues1999,berdyugina2005,berdyugina2007}. This scenario is compatible with
the currently accessible spectropolarimetric data, since two out of those seven objects do show
the presence of a magnetic field 
\citep[SDSS J111341.33+014641.7 and SDSS J133359.86+001654.8,][]{schmidt2003}.\footnote{No
spectropolarimetric measurements are available for the remaining five objects.} A serious challenge
to this hypothesis, however, is the existence of highly magnetized DQ white dwarfs with undistorted
Swan bands (GJ~1086, \citealt{berdyugina2007}; WD~1235+422, \citealt{vornanen2013}). It is
unclear how a strong magnetic field could induce a shift in some stars and not in others.

Another peculiar object is SDSS J161847.38+061155.2. The SDSS spectrum of this object appears
to show distorted Swan bands (Figure~\ref{fig:weird}), but the effective temperature derived
from the photometry (8700\,K) is a few thousand degrees hotter than any other known DQpec white dwarf. 
A visual inspection of the SDSS and Pan--STARRS images has revealed that the field is not especially
crowded so that the photometry does not seem to be contaminated by a blend of multiple objects.
SDSS J161847.38+061155.2
could possibly be an unresolved binary, although this hypothesis is unlikely given the very high surface
gravity of $\log g = 8.6$ (and thus the small effective radius) obtained from the {\it Gaia} parallax.

\section{The evolution of cool DQ/DQ\lowercase{pec} white dwarfs}
\label{sec:evolution}
\subsection{The low--$T_{\rm eff}$ portion of the two DQ sequences}
\begin{figure*}
  \centering
  \hspace{3.em}
  \includegraphics[width=1.5\columnwidth]{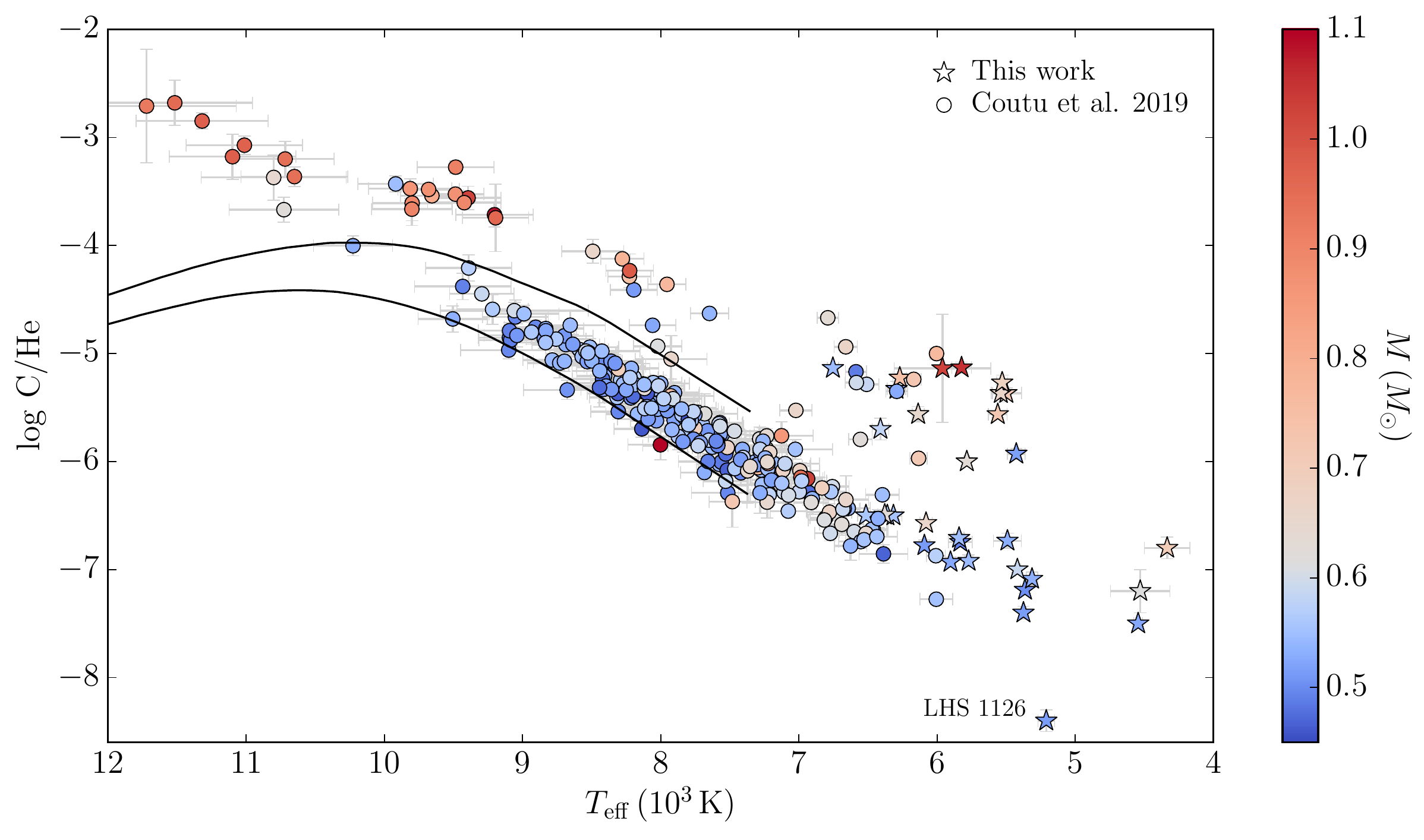}
  \caption{Carbon abundance as a function of effective temperature for the white dwarfs
    listed in Table~\ref{tab:sol} (stars) and for the DQ stars of \citet[circles]{coutu2019}.
    The solid lines correspond to the evolutionary models of \citet[]{fontaine2005} assuming
    $0.6\,M_{\odot}$ and $\log q({\rm He})=-3$ (top) or $\log q({\rm He})=-2$ (bottom). Objects
    with $M<0.45\,M_{\odot}$ were omitted from this figure as they are likely part of a binary system
    \citep{liebert2005,rebassa2011}.}
  \label{fig:teffche}
\end{figure*}

Figure~\ref{fig:teffche} shows how the carbon abundance evolves with decreasing \Teff in DQ/DQpec white dwarfs. 
This figure clearly demonstrates that the two DQ sequences previously identified at
$T_{\rm eff} \gtrsim 6500\,{\rm K}$ continue their courses down to the coolest DQpec white dwarfs.
More precisely, a large fraction of DQpecs appears to correspond to the evolved versions of the
``normal'' DQs (i.e., normal-mass objects for which the atmospheric carbon originates from the dredge-up process) 
and a smaller fraction, the more massive and carbon-rich objects (e.g., SDSS J111341.33+014641.7              
and SDSS J133359.86+001654.8),
are likely descendants of Hot DQs. We note, however, that the origin of the normal-mass white dwarfs
that lie above the main DQ sequence remains unclear.

These results, based on a detailed star-by-star analysis, imply that DQpec white dwarfs cannot 
all be descendants of Hot DQs as proposed by \cite{koester2019}.
Note that their conclusion was reached based on the high surface gravities obtained from the
{\it Gaia} parallaxes and SDSS photometry while assuming a fixed effective temperature
and carbon abundance for all DQpec white dwarfs. In hindsight, the $T_{\rm eff}=7500\,{\rm K}$ and
$\log\,{\rm C/He}=-5$ values assumed for all objects by \cite{koester2019} were unrealistic, which explains
why they found an average surface gravity of $\log g=8.625$.

\begin{figure}
  \centering
  \includegraphics[width=\columnwidth]{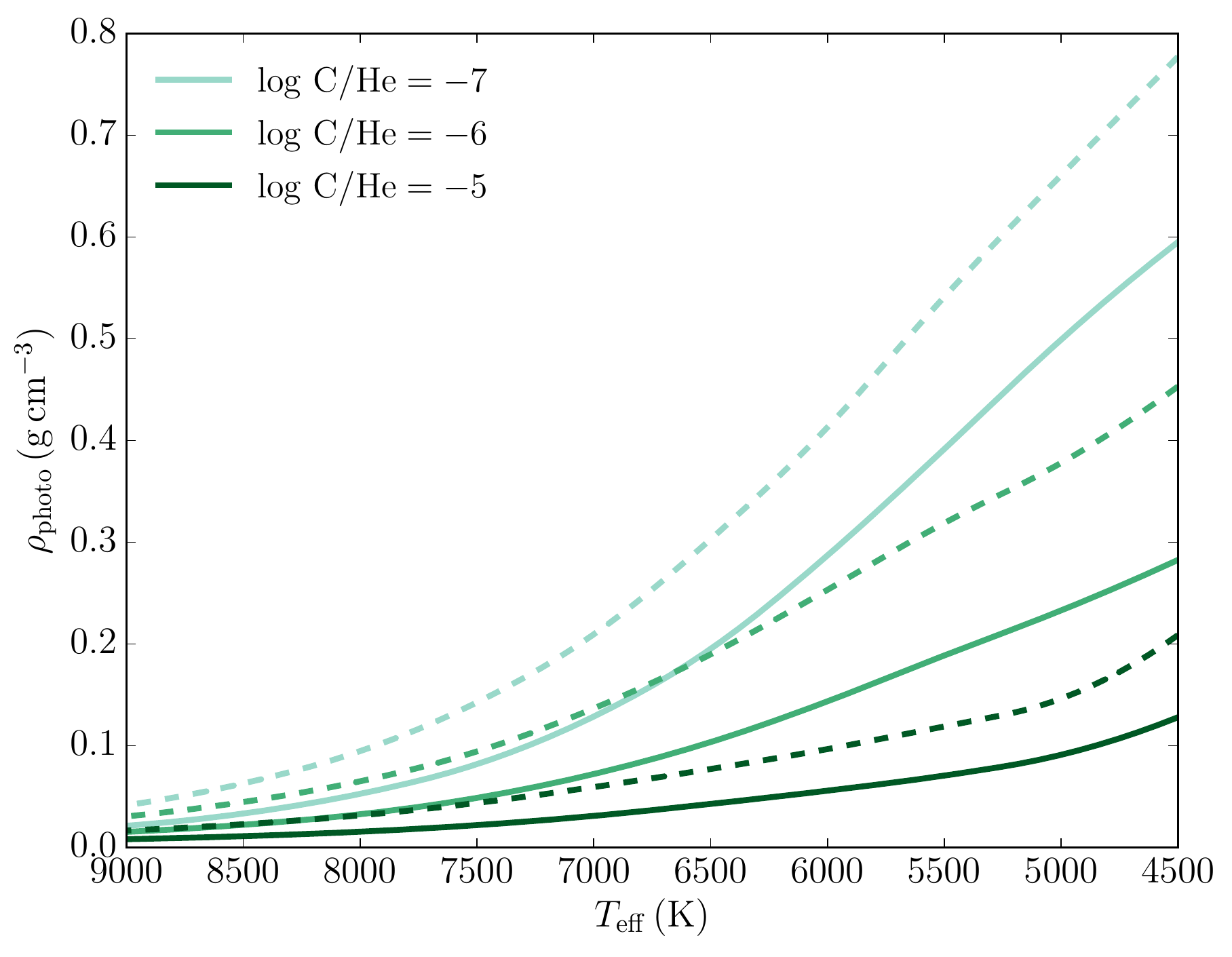}
  \caption{Density at the photosphere ($\tau_R=2/3$) of carbon-polluted white dwarfs as a function of \logche,
  \Teff and $\log g$. The solid lines are for models with $\log g=8.0$ and the dashed lines
  are for $\log g=8.5$.}
  \label{fig:rhophoto}
\end{figure}

Another argument of \cite{koester2019} to support the scenario that DQpec white dwarfs are massive
objects is that a high surface gravity can help increase the photospheric density and thus explain
the density-driven shift of the Swan bands. However, our atmosphere models clearly demonstrate that
this shift can happen in $\log g=8$ white dwarfs. A high photospheric density can be achieved as
long as the effective temperature and the carbon abundance are low enough (Figure \ref{fig:rhophoto}). 
A cool, carbon-poor atmosphere has fewer free electrons than a hot, carbon-rich atmosphere, 
which implies that He$^-$ free--free (the main opacity in the atmosphere of those objects) is less prominent.
This leads to a more transparent atmosphere and thus a photosphere that is located deeper in the star.

There is one object in Figure~\ref{fig:teffche} that is an obvious outlier from the two DQ sequences.
The very low \logche value obtained for LHS~1126 in \cite{blouin2019c} implies that it is significantly
below the main DQ sequence. However, the atmospheric parameters of this object remain highly
uncertain as current models do not allow a satisfactory fit of its spectral energy distribution (SED).
Many fits have been attempted, but none can explain all its SED from the ultraviolet to
the mid-infrared \citep{bergeron1994,wolff2002,giammichele2012,blouin2019c}. The challenge resides
in simultaneously fitting its Ly$\alpha$ red wing, distorted Swan bands and infrared flux depletion.

\subsection{The DQ$\rightarrow$DQpec transition}
\label{sec:transition}
\begin{figure}
  \centering
  \includegraphics[width=\columnwidth]{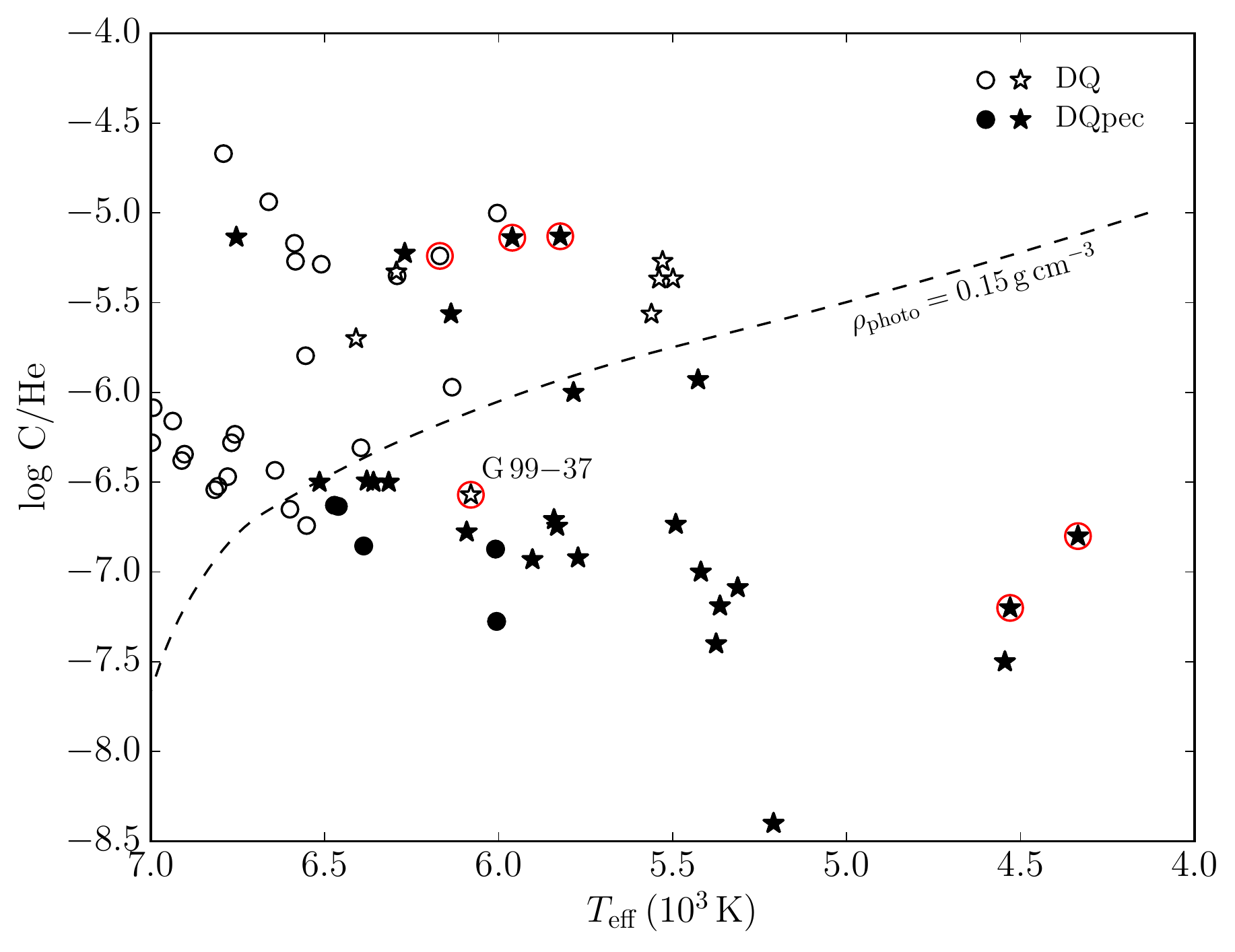}
  \caption{$T_{\rm eff}-\log\,{\rm C/He}$ diagram. The circles represent objects analysed in \citet[]{coutu2019} and
  the stars are objects analysed in this work. The filled symbols correspond to DQpec white
  dwarfs, while the empty ones are DQs. Symbols encircled in red are known magnetic white dwarfs.
  The dashed line delimits the region where the photospheric
  density exceeds 0.15\,g\,cm$^{-3}$ (assuming $\log g=8$), which marks the location of the 
  DQ$\rightarrow$DQpec transition for most objects.}
  \label{fig:transition}
\end{figure}

Another interesting aspect of the evolution of cool carbon-polluted white dwarfs is the question of where
the DQ$\rightarrow$DQpec transition occurs. Figure~\ref{fig:transition} answers this question by indicating
which objects are DQs and which ones are DQpecs in a $T_{\rm eff}-\log\,{\rm C/He}$ diagram.
For this purpose, we needed a definition of the distinction between DQs and DQpecs. We thus
performed two fits for each object, one
in which the model grid includes a shift of the Swan bands and one in which the models assume that the
Swan bands remain unperturbed by high-density effects. If the spectroscopic fit is significantly better
when the shift is included, then the star is deemed a DQpec; otherwise, it is a DQ. As in the case of Figure~\ref{fig:teffche},
objects with $M<0.45\,M_{\odot}$ are ignored. Note also that many objects of the \cite{coutu2019} sample are
not shown in this figure as their low signal-to-noise spectra did not allow us to meaningfully distinguish
between a DQ and a DQpec classification.

Figure~\ref{fig:transition} shows that the DQ$\rightarrow$DQpec transition occurs along a diagonal
line in a $T_{\rm eff}-\log\,{\rm C/He}$ diagram. Given the density-driven nature of the distortion
of the C$_2$ Swan bands and the relation between the photospheric density and \Teff and \logche 
(Figure~\ref{fig:rhophoto}), this result is not surprising. As indicated
in Figure~\ref{fig:transition}, the distortion of the Swan bands (i.e., the DQpec phenomenon) starts
to be detectable when the photospheric density (i.e., $\tau_R=2/3$) exceeds $\approx 0.15\,\textrm{g\,cm}^{-3}$.

There are however a few noteworthy exceptions to this pattern. SDSS J111341.33+014641.7, SDSS J133359.86+001654.8,
SDSS J183500.21+642917.0, SDSS J223224.00-074434.3 and SDSS J225901.16+215843.9 are DQpec stars that have
photospheric densities significantly below the $0.15\,\textrm{g\,cm}^{-3}$ threshold.\footnote{SDSS J111341.33+014641.7
and SDSS J133359.86+001654.8 are very massive white dwarfs (1.03 and 1.05\,\Msun). This implies that the
$\rho_{\rm photo}=0.15\,{\rm g}\,\textrm{cm}^{-3}$ boundary of Figure~\ref{fig:transition}, which was computed
assuming $\log g=8$, does not apply to those objects. However, even when we account for their high masses,
their photospheric densities remain below $0.15\,\textrm{g\,cm}^{-3}$.} Those five objects were already identified
in Section~\ref{sec:problems} as objects for which our models underestimate the Swan bands shift. As explained
above, it remains unclear whether this is due to our limited understanding of the behaviour of
Swan bands under high-density conditions or to the presence of strong magnetic fields that alter the structure
and opacities of those objects.

Another outlier is G\,99$-$37 (GJ 1086), which shows undistorted Swan bands despite being surrounded by DQpec 
objects in Figure~\ref{fig:transition}. This is probably explained by the presence of hydrogen in its
atmosphere (G\,99$-$37 has a CH $G$ band), which lowers its photospheric density. However, we note that
this phenomenon is still not fully understood, as the hydrogen abundance required to match the CH $G$ band 
is insufficient to inhibit the distortion of the C$_2$ Swan bands \citep[][Section 3.3.2]{blouin2019c}.

\section{Conclusions}
\label{sec:conclu}
A detailed star-by-star analysis of cool carbon-polluted white dwarfs was presented. We obtained
good spectroscopic fits for the majority of objects in our sample, including cool DQs with very
strong Swan bands and DQpec white dwarfs with strongly shifted bands. Our analysis reveals
that cool DQ/DQpec stars follow the two evolutionary sequences previously identified
at higher effective temperatures. A large fraction of objects appear to be
white dwarfs that have dredged-up carbon from their cores and a smaller fraction could be
descendants of Hot DQs. Our results imply that DQpecs represent the evolved versions
of DQ white dwarfs no matter what the origin of carbon in those stars was.

For most objects, we find that the DQ$\rightarrow$DQpec transition occurs when the photospheric density
reaches ${\approx 0.15\,\textrm{g\,cm}^{-3}}$. However, a few heavily polluted objects do not follow this trend
and display distorted Swan bands even if their photospheric densities are significantly below 
$0.15\,\textrm{g\,cm}^{-3}$. This behaviour might be due to the presence of strong magnetic fields that
can distort molecular bands. More efforts on both the observational and theoretical fronts are needed
to clarify the nature of those objects. The number of DQpec white dwarfs for which spectropolarimetric
data is available remains limited and the precise impact of strong magnetic fields on the C$_2$ Swan bands
is unclear.

\section*{Acknowledgements}
We thank the anonymous referee for providing useful suggestions that have improved the paper.
S.B. is grateful to Didier Saumon for useful discussions that have improved
the clarity of this work.

Research presented in this article was supported
by the Laboratory Directed Research and Development program of Los
Alamos National Laboratory under project number 20190624PRD2.

This work has made use of data from the European Space Agency (ESA) mission
{\it Gaia} (\url{https://www.cosmos.esa.int/gaia}), processed by the {\it Gaia}
Data Processing and Analysis Consortium (DPAC,
\url{https://www.cosmos.esa.int/web/gaia/dpac/consortium}). Funding for the DPAC
has been provided by national institutions, in particular the institutions
participating in the {\it Gaia} Multilateral Agreement.

The Pan--STARRS1 Surveys (PS1) and the PS1 public science archive have been made possible through contributions by the Institute for Astronomy, the University of Hawaii, the Pan--STARRS Project Office, the Max-Planck Society and its participating institutes, the Max Planck Institute for Astronomy, Heidelberg and the Max Planck Institute for Extraterrestrial Physics, Garching, The Johns Hopkins University, Durham University, the University of Edinburgh, the Queen's University Belfast, the Harvard-Smithsonian Center for Astrophysics, the Las Cumbres Observatory Global Telescope Network Incorporated, the National Central University of Taiwan, the Space Telescope Science Institute, the National Aeronautics and Space Administration under Grant No. NNX08AR22G issued through the Planetary Science Division of the NASA Science Mission Directorate, the National Science Foundation Grant No. AST--1238877, the University of Maryland, Eotvos Lorand University (ELTE), the Los Alamos National Laboratory, and the Gordon and Betty Moore Foundation.

\bibliographystyle{mnras}
\bibliography{references}

\bsp	
\label{lastpage}
\end{document}